# Electrically Accessible Metamagnetic Transition via a Doping-Induced Low-Energy Magnetic State in Antiferromagnetic Insulator $R$FeO$_3$

Wanting Yang[1,2], Haohuan Peng[3], Ziming Wang[1], Xiaoxuan Ma[1,2], Baojuan Kang[1], Chang Xue[1], Rongrong Jia[3], Jun-Yi Ge[1*], Jinrong Cheng[2], Shixun Cao[1*]

[1]*Materials Genome Institute, Institute of Quantum Science and Technology, International Center for Quantum and Molecular Structures, Shanghai University, Shanghai, 200444, China*

[2]*School of Material Science and Engineering, Shanghai University, Shanghai 200444, China*

[3]*Institute of Physics, Shanghai University, Shanghai, 200444, China*

Corresponding Authors: junyi_ge@t.shu.edu.cn (Jun-Yi Ge); sxcao@shu.edu.cn (Shixun Cao)

## ABSTRACT

Low-energy antiferromagnetic phase transitions offer an appealing platform for low-power spintronic functionalities, yet their direct electrical access in insulating antiferromagnets remains challenging, particularly in the low-field regime where subtle Néel vector reorientations dominate. Here, we demonstrate that targeted rare-earth-site engineering enables an electrically accessible metamagnetic transition in the insulating orthoferrite Ho$_{0.5}$Dy$_{0.5}$FeO$_3$. By combining the distinct spin-reorientation sequences of DyFeO$_3$ ($\Gamma_4 \rightarrow \Gamma_1$) and HoFeO$_3$ ($\Gamma_4 \rightarrow \Gamma_2$), Dy substitution stabilizes a dual spin-reorientation pathway $\Gamma_4 \rightarrow \Gamma_1 \rightarrow \Gamma_2$, hosting an intermediate $\Gamma_{12}$ state with a reduced energy barrier. This low-energy antiferromagnetic state can be tuned into the weak-ferromagnetic $\Gamma_4$ state under low magnetic fields. The critical field decreases with increasing temperature, providing a favorable window for functional manipulation. Both longitudinal and transverse spin Hall magnetoresistance channels exhibit clear and reproducible signatures of the metamagnetic transitions. Owing to the enhanced sensitivity of the transverse channel, additional low-field features are resolved, reflecting the projection of the Néel vector onto the spin-accumulation direction. Electrical transport measurements correlate directly with the magnetically determined phase boundaries, establishing a purely electrical access to low-energy phase transitions and to illustrate a viable pathway for exploring low-power spin dynamics in insulating oxide antiferromagnets.

## INTRODUCTION

Antiferromagnetic (AFM) materials are attracting increasing interest for next-generation spintronic technologies due to their ultrafast spin dynamics, robustness against external magnetic perturbations, and absence of stray fields[1–3]. Beyond these intrinsic advantages, a central challenge in AFM spintronics lies in the ability to electrically access and track the reorientation of low-energy barrier spin reconfigurations, particularly those occurring under weak magnetic fields. This requirement becomes especially stringent in antiferromagnetic insulators, where conventional charge-transport-based probes are unavailable. While optical and magnetic techniques have provided valuable insights into AFM phase transitions, a purely electrical probe capable of resolving low-energy spin reconfiguration processes remains highly desirable for both fundamental studies and device-oriented applications. Spin Hall magnetoresistance (SMR) has recently emerged as a powerful probe for such magnetic reconfigurations[4–7]. Unlike conventional transport measurements, SMR provides an interface-mediated electrical access to magnetic order in insulating antiferromagnets via spin-current reflection at magnetic insulator/metal interfaces. However, its applicability to noncollinear antiferromagnets and, in particular, to low-energy metamagnetic transitions involving intermediate spin states has remained largely unexplored[8,9]. In such systems, the Néel-vector (***n***) undergoes subtle and often continuous reorientation within a shallow anisotropy landscape, posing stringent requirements on the sensitivity of electrical detection schemes.

Rare-earth orthoferrites $R$FeO$_3$ ($R$ = La-Lu, Y) provide a compelling platform for investigating magnetic phase transition and tunable magnetic anisotropy governed by competing exchange interactions[10,11]. $R$FeO$_3$ crystallizes in an orthorhombic *Pbnm* perovskite structure, where Dzyaloshinskii-Moriya interaction (DMI) arising from octahedral rotations induces a weak ferromagnetic (FM) canting of the $Fe^{3+}$ spin[12–14]. The strong Fe-Fe superexchange stabilizes a *G*-type AFM order with a high Néel temperature of $Fe^{3+}$ ($T_{N1} \approx$ 600-750 K) and [15], while the $R^{3+}$ sublattice typically remains weakly ordered ($T_{N2} <$ 10 K) and is strongly influenced by the $Fe^{3+}$ molecular field at higher temperatures[16]. The resulting 4*f*-3*d* exchange interaction plays a crucial role in determining the magnetic anisotropy landscape and stabilizing competing spin configurations of the $Fe^{3+}$ sublattice ($\Gamma_1$ ($A_x$, $G_y$, $C_z$), $\Gamma_2$ ($F_x$, $C_y$, $G_z$), or $\Gamma_4$ ($G_x$, $A_y$, $F_z$))[17–19]. Therefore, $RFeO_3$ exhibit temperature-driven spin reorientation transition (SRT) and field-driven metamagnetic transitions, during which ***n*** rotates between nearly degenerate easy axes. Depending on the $R^{3+}$, these transitions may proceed via distinct pathways. SRT can be commonly classified as $\Gamma_4 \rightarrow \Gamma_2$ (as in $HoFeO_3$[20]) and $\Gamma_4 \rightarrow \Gamma_1$ (as in $DyFeO_3$[21]), thereby providing a natural basis for *R*-site engineering. In addition, the $R^{3+}$ and $Fe^{3+}$ spin can be influenced to flip with the temperature/magnetic field change (spin switching, SSW)[22]. The type-I SSW occurs when both the $R^{3+}$ and $Fe^{3+}$ spins flip simultaneously[23–25], while the type-II SSW occurs if only the $R^{3+}$ spin flips[26,27], which further enriches the accessible magnetic states. Despite this tunability, an electrical probe capable of resolving low-energy metamagnetic transitions associated with intermediate states is still lacking.

In this work, we address this challenge by performing SMR-based transport and magnetization measurements on $Ho_{0.5}Dy_{0.5}FeO_3$ single crystals and their Pt heterostructures. By tailoring the *R* sublattice, we realize a temperature-driven dual SRT involving an intermediate $\Gamma_{12}$ state with a comparatively low energy barrier. This intermediate AFM configuration further enables a field-induced metamagnetic transition into the high-field $\Gamma_4$ state. Both longitudinal and transverse SMR channels provide clear electrical signatures of these metamagnetic phase transition, demonstrating a pronounced sensitivity of interfacial spin-current interactions to Néel vector reorientation. By correlating SMR with magnetometry, we refine the magnetic phase boundaries and reveal a field-driven collapse of the $\Gamma_{12}$ state with increasing temperature. Our results establish SMR as a sensitive and purely electrical probe of low-energy phase transition in antiferromagnetic insulators, thereby offering new opportunities for exploring spin transport in non-collinear antiferromagnetic systems.

## EXPERIMENTAL DETAILS

**Synthesis and Crystal Growth**: Polycrystalline $Ho_{0.5}Dy_{0.5}FeO_3$ was synthesized via solid-state reaction from high-purity $Ho_2O_3$ (99.99%), $Dy_2O_3$ (99.99%) and $Fe_2O_3$ (99.95%). The stoichiometrically mixed powders were sintered at 1200 °C for 1000 minutes to ensure complete reaction. Single crystal was subsequently grown in air atmosphere using an optical floating zone furnace (FZ-T-10000-H-VI-P-SH, Crystal Systems Corp.) with a growth rate of 3 mm/h.

**Structural Characterization**: Portions of the as-grown $Ho_{0.5}Dy_{0.5}FeO_3$ single crystal were ground into fine powders and subjected to powder X-ray diffraction (XRD, Bruker D2 PHASER). Crystal orientation was determined by back-reflection Laue photography (TrySE. Co, Ltd.). Oriented samples were then precision-cut into a cube (about $2.5 \times 2.5 \times 2.5$ mm$^3$) and a thin slice (about $4.0 \times 4.0 \times 0.7$ mm$^3$).

**Device Fabrication for SMR**: The single-crystal slice was mechanically polished to achieve sub-nanometer-scale surface roughness (see Supporting Information S1). A Hall bar device pattern was then designed using Klayout software and fabricated using a maskless lithography system (MicroWriter ML3). A Pt film with a thickness of 7 nm was 10 nm subsequently deposited by electron-beam evaporation (VZZS-550) and patterned into a Hall bar geometry with a channel width of 15 μm and a probe spacing of 70 μm.

**Magnetic and Electrical Measurements**: Magnetic characterization was performed using a Vibrating Sample Magnetometer (VSM) equipped in a Physical Property Measurement System (PPMS-14T, Quantum Design Inc.). Temperature- and magnetic field-dependent magnetization (*M-T*, *M-H*) curves were recorded in zero-field-cooled warming (ZFC) or field-cooled cooling (FCC) modes, with reproducible results. Electrical transport measurements were carried out using a Keithley 2400 current source and a Keithley 2182A nanovoltmeter, integrated with the PPMS via a customized resistor interface box to enable high-precision transport characterization under controlled temperature and magnetic field.

## RESULT AND DISCUSS

**Structural Characterization**: The powder X-ray diffraction (XRD) spectra of $Ho_{0.5}Dy_{0.5}FeO_3$ and the corresponding Rietveld refinement are shown in Figure 1a. The calculated profile reproduces the experimental data well, yielding a reliability factor $R_{wp}$ = 10.2%. No impurity reflections are detected, confirming the high phase purity of the crystal. The refinement verifies that the compound crystallizes in the orthorhombic perovskite structure (space group *Pbnm*) with lattice parameters $a$ = 5.2959(4) Å, $b$ = 5.5989(1) Å, and $c$ = 7.6201(3) Å. The crystal structure visualized using VESTA is shown in Figure1b and 1c for projections along the $y$ and $z$ directions, respectively. The complementary views explicitly reveal the cooperative tilting and rotation of the $FeO_6$ octahedra[28]. Such octahedral distortions break the local inversion symmetry at the Fe-O-Fe bonds and provide the structural origin of the DMI, which underlies the weak ferromagnetic canting and the spin-reorientation behavior discussed below.

The as-grown high quality $Ho_{0.5}Dy_{0.5}FeO_3$ signal crystal exhibits a metallic luster and a rod-like morphology with a length of approximately 53 mm and a diameter of about 5 mm (Figure1d). The oriented single-crystal XRD scans were employed to determine the crystallographic orientation. The patterns recorded along the $a$, $b$, and $c$-axes in (Figure 1e-g) exhibit exclusively ($h$00) ($h$ = 2, 4), (0$k$0) ($k$ = 2), and (00$l$) ($l$ = 2, 4, 6), respectively. The corresponding Laue back-reflection patterns in the insets show sharp and well-defined diffraction spots, demonstrating the high crystallinity of the single crystal. The angular deviations extracted from Laue indexing are below 0.05°, ensuring the precision of the crystallographic alignment for the magnetization and SMR measurements.

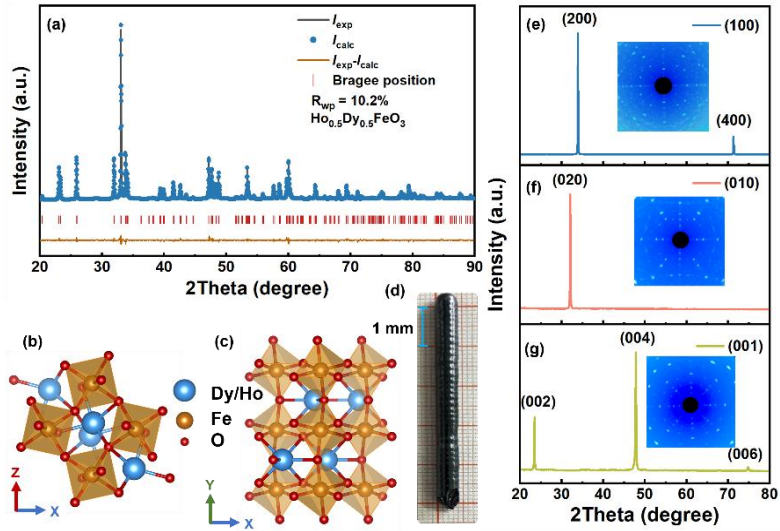

Figure 1. Structural characterization of $Ho_{0.5}Dy_{0.5}FeO_3$. (a) Rietveld-refined X-ray diffraction pattern, showing high agreement between the experimental ($I_{exp}$) and calculated ($I_{calc}$) profiles and confirming a single-phase orthorhombic perovskite structure. (b-c) Crystal structure viewed along $y$- and $z$-perspectives, respectively; Dy/Ho, Fe, and O ions are shown as blue, orange, and red spheres, respectively, with $FeO_6$ octahedra highlighted in orange. Lattice parameters are extracted from the Rietveld refinement in (a). (d) Picture of the signal crystal with a scale bar of 1 mm. (e-g) X-ray diffraction and corresponding Laue back-reflection patterns (insets) recorded along three orthogonal axes, demonstrating high crystallinity and precise orientation.

**Magnetic measurements**: Figure 2 presents the temperature-dependent magnetization measured along the crystallographic $a$- and $c$-axes for $HoFeO_3$, $DyFeO_3$ and $Ho_{0.5}Dy_{0.5}FeO_3$ single crystals to illustrate the evolution of SRT behavior from the parent compounds to the mixed rare-earth system. This comparative presentation highlights the distinct SRT pathways inherited from the parent compounds and elucidates the $R$-Fe exchange interplay in the mixed-rare-earth system. For $HoFeO_3$ Figure 2a), a characteristic SRT of $\Gamma_4$ ($G_x$, $F_z$)→$\Gamma_2$ ($F_x$, $G_z$) state is observed. In contrast, $DyFeO_3$ (Figure 2b) exhibits a distinct $\Gamma_4$ ($G_x$, $F_z$)→$\Gamma_1$ ($A_x$, $C_z$) transition, manifested by the abrupt suppression of the $c$-axis magnetization as the weak-ferromagnetic canting vanishes.

Against this comparative backdrop, the temperature evolution of $Ho_{0.5}Dy_{0.5}FeO_3$ (Figure 2c) reveals a modified SRT behaviour. At high temperatures ($T > 51$ K), $Ho_{0.5}Dy_{0.5}FeO_3$ stabilizes in the $\Gamma_4$ ($G_x$, $F_z$) state, where **n** predominantly aligns along the $a$-axis and a weak net magnetic moment **m** appears along the $c$-axis due to DMI-induced canting. Upon cooling, the magnetization gradually increases, reflecting the progressive evolution of the $Fe^{3+}$ magnetic order and the accompanying evolution of the canting angle. A sharp Dy-driven $\Gamma_4$ ($G_x$, $F_z$)→$\Gamma_1$ ($A_x$, $C_z$) SRT occurs at 51 K, manifested as an abrupt suppression of the c-axis magnetization to almost zero, while the a-axis magnetization remains negligible. However, the $\Gamma_1$ phase does not represent the ground state of $Ho_{0.5}Dy_{0.5}FeO_3$. With further cooling, the magnetization along the a-axis emerges gradually, indicating a continuous rotation of *n* towards the c-axis. The system enters an intermediate state $\Gamma_{12}$ between 51 K and 23 K, in which competing R-Fe exchange interactions nearly balance the magnetic anisotropies along the a- and c-axes and resulting a comparatively low magnetic energy barrier. At 23 K, a second SRT from $\Gamma_1$ ($A_x$, $C_z$)→$\Gamma_2$ ($F_x$, $G_z$) completes. The observed dual SRT sequence $\Gamma_4$→$\Gamma_1$→$\Gamma_2$ in $Ho_{0.5}Dy_{0.5}FeO_3$ reflects a delicate competition between the anisotropic symmetric and antisymmetric exchange interactions mediated by the mixed $R^{3+}$ sublattice. This competition gives rise to the intermediate $\Gamma_{12}$ state, providing a platform for exploring low-energy-barrier antiferromagnetic phase transitions.

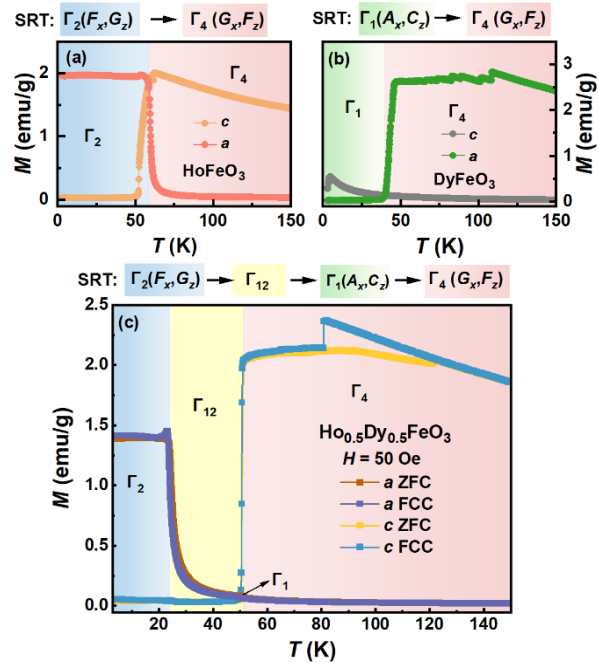

Figure 2. Temperature-dependent magnetization measured under $H$ = 50 Oe along the crystallographic a- and c-axes using zero-field-cooled and field-cooled protocols for (a) $HoFeO_3$, (b) $DyFeO_3$ and (c) $Ho_{0.5}Dy_{0.5}FeO_3$ single crystals. The shaded green, yellow, blue, and pink regions denote $\Gamma_1$, $\Gamma_{12}$, $\Gamma_2$, and $\Gamma_4$ magnetic states, respectively.

To clarify the field-dependent magnetic behavior of $Ho_{0.5}Dy_{0.5}FeO_3$, the magnetic field-temperature phase diagram is summarized in Figure 3a, together with representative magnetization curves along the *a*- and *c*-axes shown in Figure 3b-g. Distinct magnetic phases are identified and highlighted by shaded regions, with schematic spin configurations overlaid to visualize the orientation of the $Fe^{3+}$ spins in each phase. Below the SRT region (Figure 3b,c), *a*-axis magnetization saturates rapidly at high fields, while *c*-axis magnetization responses linearly. This behaviour indicates the weak ferromagnetism along the *a*-axis and antiferromagnetism along *c*-axis, corresponding to the $\Gamma_2$ state. Thus, the spins of $Fe^{3+}$ align along the *x*-direction with a slight canting along the *x*-direction, as shown in the inset of Figure 1a. Within the SRT region, the weak ferromagnetic component along the *a*-axis is partially suppressed, and the corresponding *M-H* curves display a reduced saturation moment and enhanced linear contribution (Figure 3d,e). The *c*-axis magnetization remains nearly linear with negligible remanence and coercivity under low fields, consistent with the intermediate $\Gamma_{12}$ state. In this state, the spin orientation deviates from the principal crystallographic axes and lie in the *yz*-plane with the reduced ***m*** along the *x*-direction. Above the SRT region (Figure 3f,g), the magnetic anisotropy reverses as the *c*-axis develops a weak ferromagnetic component, while the *a*-axis behaves antiferromagnetically. This phenomenon corresponds to the $\Gamma_4$ phase in the phase diagram, where ***n*** already rotates to the *x*-direction and ***m*** lies along the *z*-direction. No signatures of the $\Gamma_1$ phase are observed in the *M-H* curves, supporting its transient nature during SRT. The complementary perspectives of magnetic structures can be more clearly observed in the supplementary materials, revealing the rotation of $Fe^{3+}$ spins as well as the orientations of ***n*** and ***m***.

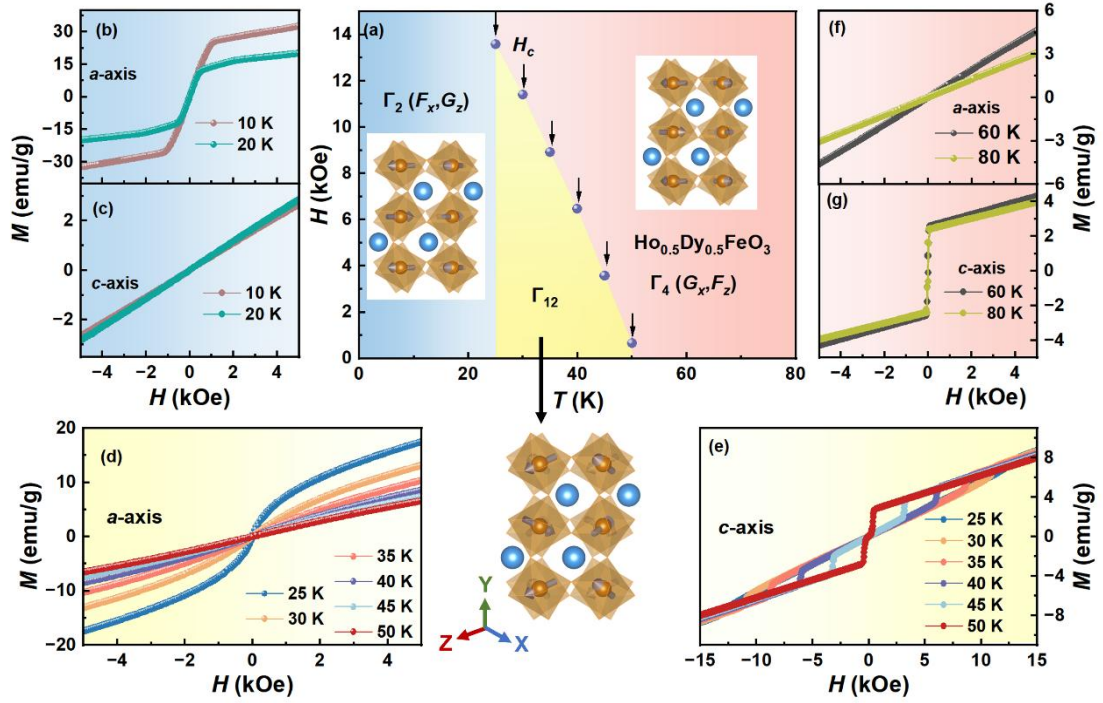

Figure 3. Magnetic phase diagram and field-dependent magnetization (*M-H*) of $Ho_{0.5}Dy_{0.5}FeO_3$ single crystal. (a) Magnetic field-temperature phase diagram, where shaded blue, yellow, and pink regions indicate $\Gamma_2$, $\Gamma_{12}$, and $\Gamma_4$ magnetic configurations, respectively. (b-c) Respective *M-H* curves along *a*- and *c*-axes in $\Gamma_2$ state. (d-e) *M-H* curves along *a*- and *c*-axes in the intermediate $\Gamma_2$ state. (f-g) *M-H* curves along *a*- and *c*-axes in $\Gamma_4$ state.

A pronounced field-induced metamagnetic phase transition is observed within the SRT range, where the Fe sublattice adopts the intermediate $\Gamma_{12}$ state, as shown in Figure 3e. This behaviour can be captured by the simplified free-energy expression:

$$E(\theta) = K_{\text{eff}}\sin^2\theta - MH\cos(\theta - \theta_H), \tag{1}$$

which highlights the competition between the magnetic anisotropy energy ($K_{\text{eff}}\sin^2\theta$) and Zeeman energy ($MH\cos(\theta - \theta_H)$). In the $\Gamma_{12}$ state, the effective anisotropy constant $K_{\text{eff}}$ is strongly suppressed, and thus the anisotropic energy between the easy and hard axes is strongly reduced. As a result, even a moderate external magnetic field can overcome the residual anisotropy and continuously rotate the ***m*** of $Fe^{3+}$ toward the *c*-axis ($\Gamma_4$ state in Figure 3a). This field-driven spin reorientation gives rise to a sharp yet continuous increase in magnetization, manifesting as a metamagnetic phase transition. The low-energy, field-tunable $\Gamma_{12}$ state provides a microscopic basis for the electrically detectable magnetic state switching behavior discussed below.

**SMR measurements**: The field-induced metamagnetic transition in

$Ho_{0.5}Dy_{0.5}FeO_3$ involves a rearrangement of the underlying $Fe^{3+}$ spin structure within a low-energy anisotropy landscape. Such a reconfiguration is expected not only to modify the macroscopic magnetization, but also to strongly affect interfacial spin-current transport, rendering the magnetic state switching directly accessible through (SMR). To probe this effect, a Pt Hall bar was fabricated on the *c*-oriented $Ho_{0.5}Dy_{0.5}FeO_3$ surface, with the transport channel aligned along the orthorhombic *a*-axis, as schematically illustrated in Figure 4a, b. In this geometry, the out-of-plane *z*-axis coincides with the crystallographic *c*-axis. A dc current density $J = 5 \times 10^8$ $Am^{-2}$ was applied along the *x*-direction in the Pt layer, generating, via the spin Hall effect (SHE), a spin accumulation $\mu_s$ polarized along *y* at the Pt/$Ho_{0.5}Dy_{0.5}FeO_3$ interface. By sweeping an external magnetic field along z, both the longitudinal and transverse resistivities of the Pt bar were simultaneously recorded, yielding the longitudinal SMR (LSMR) and transverse SMR (TSMR) responses. Details of device fabrication and background-signal subtraction are provided in Supporting Information S3.

Figure 4a displays the field dependence of the LSMR at representative temperatures within the SRT region. At low fields, the LSMR exhibits a pronounced magnitude, where *n* in the intermediate $\Gamma_{12}$ state is misaligned with $\mu_s$, leading to efficient spin-current absorption. With increasing field, *n* continuously rotates toward the $\Gamma_4$ configuration, reducing its projection onto $\mu_s$ and thereby suppressing the SMR amplitude. Once the system enters the high-field $\Gamma_4$ state, the SMR contribution collapses, and the remaining background signal dominates by ordinary magnetoresistance, which has been subtracted. The reduction of LSMR reflects that *n* rotates away from the *yz*-plane towards the *x*-direction, corresponding to the magnetic state switching from $\Gamma_{12}$ state to $\Gamma_4$ state. Additional longitudinal SMR measurements with *H*//*z* and *I*//*y* confirms that the observed SMR response is primarily governed by the *c*-axis spin configuration rather than the in-plane current direction (see Supporting Information S4)."

The corresponding TSMR, shown in Figure 4b provides an even more polarization-sensitive probe of the spin configuration. After removing the ordinary Hall contribution using a symmetry-based procedure (Supporting Information S3), the extracted TSMR characteristic a distinct low-field negative dip of the $\Gamma_{12}$ state, in which *n* retains a substantial projection along $\mu_s$. Upon increasing the field, the TSMR exhibits a sharp rise followed by a peak, marking the magnetic state switching into the $\Gamma_4$ phase. The fact that both the LSMR and TSMR channels capture this switching provides unambiguous electrical evidence of the metamagnetic phase boundary.

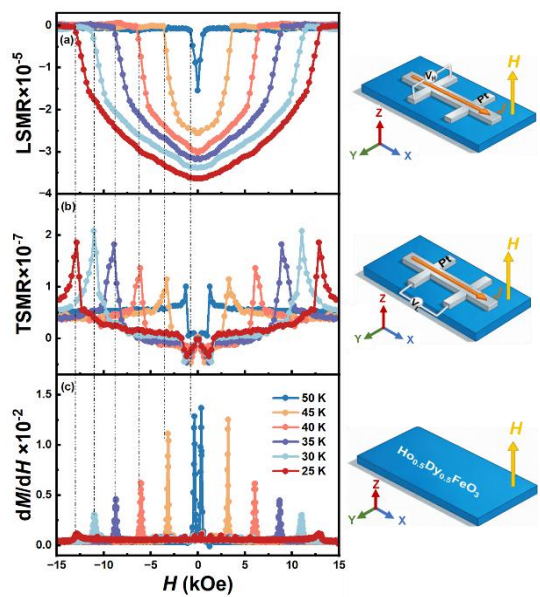

Figure 4. Spin Hall magnetoresistance and metamagnetic transition in $Ho_{0.5}Dy_{0.5}FeO_3$. (a)Longitudinal and(b)transverse spin Hall magnetoresistance measured as a function of magnetic field with $H // z$ and $j // x$; schematics of the corresponding device geometries are shown alongside each panel. Contributions from ordinary magnetoresistance and ordinary Hall effect backgrounds are subtracted. (c) Field derivative d$M$/d$H$ extracted from the *c*-axis *M-H* curves in Figure 3e, where sharp peaks identify the critical fields of the metamagnetic transition; the schematic illustrates the magnetization measurement configuration.

To establish a direct magnetic reference for the electrically observed transition, Figure 4c shows d$M$/d$H$ derived from the *c*-axis *M-H* curves. Sharp peaks in d$M$/d$H$ define the critical fields $H_c$ of the metamagnetic transition. Moreover, $H_c$ extracted independently from d$M$/d$H$, LSMR, and TSMR, which represent responses of fundamentally different physical quantities, coincide within experimental uncertainty. The critical field in magnetization reflects the nonlinear evolution of the weak ferromagnetic canting, whereas that in LSMR originates from abrupt changes in the interfacial spin-current absorption efficiency, governed by the relative orientation between *n* and $\mu_s$. The TSMR channel, being even more polarization selective, responds sensitively to the detailed rotation path of *n*. The consistency across these channels establishes the metamagnetic transition as a robust, electrically addressable magnetic state switching, directly relevant for spin-transport functionality in insulating antiferromagnets.

.

## CONCLUSION

By engineering the $R$ sublattice through partial Dy substitution at the Ho site, we realize a $Ho_{0.5}Dy_{0.5}FeO_3$ single crystal that exhibits a temperature-driven dual SRT $\Gamma_4 \rightarrow \Gamma_1 \rightarrow \Gamma_2$ accompanied by an intermediate $\Gamma_{12}$ configuration. The mixed-$R^{3+}$ design reconstructs the anisotropy landscape through competing $4f$-$3d$ exchange interactions, giving rise to this low-energy $\Gamma_{12}$ state absent in the parent compounds. Owing to its reduced energy barrier, the $\Gamma_{12}$ state is highly susceptible to external magnetic fields and undergoes a field-induced metamagnetic transition into the $\Gamma_4$ state via a collective, free-energy-driven reconfiguration of ***n***. The resulting $H$-$T$ phase diagram reveals a systematic reduction of $H_c$ with increasing temperature, reflecting the gradual weakening of the $R$-Fe exchange fields that stabilize the low-field spin configuration. By correlating magnetization with LSMR and TSMR measurements, we demonstrate that the metamagnetic transition is not only magnetically identifiable but is directly imprinted in spin-transport responses. The reorientation of ***n*** across the metamagnetic transition produces a pronounced and reproducible magnetic-state switching behavior in both LSMR and TSMR channels, reflecting an active modulation of interfacial spin-current absorption and polarization. The high agreement among $H_c$ extracted from d$M$/d$H$, LSMR, and TSMR establishes SMR as a polarization-selective electrical channel for magnetic state switching in insulating antiferromagnets. These findings reveal the strong coupling between antiferromagnetic spin structures and interfacial spin-current processes in $R$FeO$_3$ and highlight that low-energy metamagnetic transitions can be actively harnessed in SMR measurements, offering a viable pathway toward low-power antiferromagnetic spintronic functionalities.